\documentclass[preprint,aps,prd,showpacs,nofootinbib]{revtex4}
\usepackage{mathrsfs}
\usepackage[mathscr]{euscript}
\usepackage{float,epsfig}
\usepackage{dcolumn}
\usepackage{bm}
\usepackage{graphicx}
\usepackage{dcolumn}
\usepackage{amsmath,amssymb,amsthm}
\usepackage[colorlinks=true,linkcolor=blue]{hyperref}
\usepackage{booktabs}

\newcommand{\bea}{\begin{eqnarray}}
\newcommand{\eea}{\end{eqnarray}}
\newcommand{\beq}{\begin{equation}}
\newcommand{\eeq}{\end{equation}}
\newcommand{\nn}{\nonumber}
\def\/{\over}
\newcommand{\bra}[1]{\langle#1|}
\newcommand{\ket}[1]{|#1\rangle}

\begin{document}
\title{Dynamics of an elementary quantum system outside a radiating Schwarzschild black hole}
\author{  Jiawei Hu,$^{1}$ Wenting Zhou$^{1}$ and Hongwei Yu$^{1,2, }$\footnote{Corresponding author} }
\affiliation{$^1$ Institute of Physics and Key Laboratory of Low
Dimensional Quantum Structures and Quantum
Control of Ministry of Education,\\
Hunan Normal University, Changsha, Hunan 410081, China \\
$^2$ Center for Nonlinear Science and Department of Physics, Ningbo
University, Ningbo, Zhejiang 315211, China}


\begin{abstract}

We study, in the framework of open quantum systems, the dynamics of a radially polarizable two-level atom in multipolar coupling to fluctuating vacuum electromagnetic fields  which is placed at a fixed radial distance outside a radiating Schwarzschild black hole, and  analyze the transition rates between atomic energy levels and the steady state the atom is driven to. We find that the atom always thermalizes toward a steady state at an  effective temperature between zero and the Hawking temperature of the black hole. Remarkably, the thermalization  temperature depends on the transition frequency of the atom, so that atoms with different transition frequencies essentially thermalize to different temperatures. This counter-intuitive behavior is however in close analogy to what happens for a two-level atom in a stationary environment out of thermal equilibrium near a dielectric body of certain geometry and dielectric permittivity. Our results thereby suggest in principle a possible analogue system, using engineered materials with certain desired dielectric properties to verify features of Hawking radiation  in tabletop experiments.
\end{abstract}

\pacs{ 03.65.Yz, 04.70.Dy, 04.62.+v, 05.70.Ln}

\maketitle

\baselineskip=16pt

In 1974, Hawking showed, in the framework of quantum field theory in curved spacetime, that a black hole is not completely black, but emits particles in a blackbody spectrum via a quantum instability~\cite{hk}. This amazing discovery has attracted widespread attention and since then extensive work has been done trying to re-derive and understand it in a variety of different physical contexts~\cite{hk,Gibbons,Parikh,SC,Robinson,hkur,yu1,yu2,hu1,hu2}. On one hand, the Hawking effect has been considered as a
“Rosetta stone” to relate quantum theory, general relativity, and thermodynamics, and it is
expected to be an indispensable part of a yet-to-be-found full theory of quantum gravity; on the other hand, a direct experimental verification of it,  as well as other quantum effects unique to curved spacetime which arise as a consequence of the combination of general relativity and quantum theory, still remains elusive.  Yet, an indisputable detection of this amazing effect has a far-reaching impact in many areas of physics such as astrophysics~\cite{Ast}, cosmology~\cite{Cosm}, and string theory that intends to unite everything~\cite{String}  and is becoming one of the main experimental challenges of our time.  As a result,  there have been, in recent years,  many attempts to observe them in analogue systems~\footnote{ See also attempts to find, instead, corrections to quantum effects already existing in flat spacetime caused by the spacetime curvature~\cite{SCILS,hu2} as opposed to those which are unique to curved spacetime.}~\cite{unruh81,analog,Garay00,Philbin08,Nation09}, such as the Bose-Einstein condensate~\cite{Garay00}, optical fibers~\cite{Philbin08}, and superconducting transmission line~\cite{Nation09}. Such efforts, primarily driven by our curiosity and desire to meet the challenge of experimentally observing the Hawking radiation, are also expected to shed light on some unanswered questions associated
with Hawking's original calculation itself, such as the trans-Planckian problem~\cite{TransP}.

The purpose of  this paper is to study the dynamics of an elementary two-level quantum system, e.g., a two-level atom, coupled to fluctuating  vacuum electromagnetic fields outside a radiating Schwarzschild black hole, and to compare it with a situation where the two-level atom is placed near a dielectric body at finite temperature, so as to suggest a possible electromagnetic  system where features of the Hawking radiation might be experimentally verified via the dynamical behaviors of the two-level atom. Here, let us note that a two-level system, despite its simplicity,  has been widely used as a prototype to understand and predict many QED phenomena, such as the occurrence of finite lifetimes, the appearance of coherences, Lamb shift and Rabi oscillations in atom-photon interactions~\cite{Cohen}, and it has also become increasingly important in the arena of quantum optics~\cite{You} and quantum information~\cite{QI}. Moreover, two-level quantum systems  have gone from toy models used for an easy grasp of basic features of the theories being studied to an experimental reality that can be implemented in actual experiments  due to rapid progress in artificial atoms made from superconducting circuits in recent years~\cite{SuperC}. 
In fact, interesting experiments using superconducting circuits to test fundamental quantum effects, such as the Lamb shift~\cite{SuperLamb} and the dynamical Casimir effect~\cite{dce1,dce2} have been reported.

The dynamical evolution of the atom is strongly connected with quantum decoherence and dissipation induced by the environment~\cite{open,dissipative}, which is taken as a bath of fluctuating electromagnetic fields outside a radiating black hole in the present paper, and so it  can be dealt with in the framework of open quantum systems by studying the reduced density matrix. One expects that the transition rates between energy levels and the thermalization process of the atom will be influenced by the presence of the Hawking radiation and the spacetime curvature which backscatters the fluctuating electromagnetic fields the atom is coupled to. We will examine how the transition rates or lifetimes are modified outside a black hole and what effective temperature the thermalization brings the atom to. We are particularly interested in finding out how this compares to what happens to a two-level atom near a dielectric body at finite temperature.


Our model is a radially polarizable two-level atom placed at a radial distance $r$ from a radiating Schwarzschild black hole which is in multipolar coupling to the fluctuating vacuum electromagnetic fields.
A black hole that is radiating into empty space is, as is well-known, best characterized by the Unruh vacuum which was first defined using a scalar field~\cite{Unruh} and  recently the definition has been extended to the electromagnetic case~\cite{zhou}  based on the Gupta-Bleuler quantization of free electromagnetic fields in a static spherically symmetric spacetime of arbitrary dimension in a modified Feynman gauge~\cite{Crispino}. So, the Hamiltonian of the total system (the atom plus a bath of fluctuating electromagnetic fields in the Unruh vacuum) takes the form
$H=H_A+H_B+H_I\;.$
Here $H_A$ is the Hamiltonian of the two-level atom, which, in general, can be written as
$H_A=\sum_{n=1}^2\omega_n\ket{n}\bra{n}\;.$
where $\ket{1}$ and $\ket{2}$ represent the ground and excited states, and $\omega_0=\omega_2-\omega_1$ is the energy level spacing of the atom. $H_B$ is the Hamiltonian of the environment, in our case, a bath of fluctuating electromagnetic fields in the Unruh vacuum. The explicit form of $H_B$ is not needed here. In the multipolar coupling scheme~\cite{CPP95}, the interaction Hamiltonian $H_I$ takes the form
$H_I(\tau)=-\textbf{D} \cdot \textbf{E}(x(\tau))\;,$
in which $\textbf{D}$ is the  electric dipole moment of the atom, and ${\bf E}(x)$ the electric field strength. The spacetime metric of a Schwarzschild black hole reads
$ds^2=(1-2M/r)\,dt^2-(1-2M/r)^{-1}dr^2-r^2(d\theta^2+\sin^2\theta\,d\phi^2)\;.$

Initially, the state of the whole system is characterized by the total density matrix $\rho_\text{tot}=\rho(0) \otimes \rho_B$, in which $\rho(0)$ is the initial reduced density matrix of the atom, and $\rho_B$ describes the state of the environment.
The evolution of the total density matrix $\rho_\text{tot}$ in the frame of the atom satisfies the quantum Liouville equation
\begin{equation}
\frac{d\rho_\text{tot}(\tau)}{d\tau}=-i\,[H(\tau),\rho_\text{tot}(\tau)]\;,
\end{equation}
We define $A_{i}(\omega)= \sum_{\epsilon'-\epsilon=\omega}\Pi(\epsilon)D_i\Pi(\epsilon')$, where the operator $\Pi(\epsilon)$ denotes the projection onto the eigenspace belonging to the eigenvalue $\epsilon$ of $H_A$ and $i\in\{r, \theta, \phi\}$,  the Fourier transform ${\cal G}_{ij}(\omega)$ of
the field correlation function ${\langle}E_i(s)E_j(0)\rangle$,
\begin{equation}\label{}
{\cal G}_{ij}(\omega)=
 \int^\infty_{-\infty}ds\,e^{i\omega{s}}{\langle}E_i(s)E_j(0)\rangle\;,
\end{equation}
and the one-side Fourier transform $\gamma_{ij}(\omega)$
\begin{equation}
\gamma_{ij}(\omega)=
 \int^\infty_0dse^{i\omega{s}}{\langle}E_i(s)E_j(0)\rangle\;.
\end{equation}
One can show that the reduced density matrix $\rho$ governing the dynamics of the atom  which can be  obtained  by tracing over the field degrees of freedom takes,  in the limit of weak coupling, the following form  \cite{Lindblad,open}
\bea
\frac{d}{d\tau}\rho(\tau)
&=&-i\big[H_A+H_{LS},\rho(\tau)\big]
 +\Gamma(-\omega_0)\bigg(\rho_{11}\ket{2}\bra{2}-\frac{1}{2}\{\ket{1}\bra{1},\rho(\tau)\}\bigg)\nn\\
&&+\Gamma(\omega_0)\bigg(\rho_{22}\ket{1}\bra{1}-\frac{1}{2}\{\ket{2}\bra{2},\rho(\tau)\}\bigg),
\eea
where the Lamb shift Hamiltonian
\begin{equation}\label{HLS}
H_{LS}=\sum_\omega\sum_{i,j}
       i\biggl(\frac{1}{2}\,{\cal G}_{ij}(\omega)-\gamma_{ij}(\omega)\biggr)A^\dag_i(\omega)A_j(\omega)\;,
\end{equation}
$\rho_{ij}=\bra{i}\rho\ket{j}$, and $\Gamma(\omega_0)$ and $\Gamma(-\omega_0)$ are the spontaneous emission and excitation rates, which, respectively , take the form
\beq\label{rate}
\begin{aligned}
\Gamma(\omega_0)&=\sum_{i,j}
{\cal G}_{ij}(\omega_0)\bra{2}D_i\ket{1}\bra{1}D_j\ket{2}\;,\\
\Gamma(-\omega_0)&=\sum_{i,j}
{\cal G}_{ij}(-\omega_0)\bra{1}D_i\ket{2}\bra{2}D_j\ket{1}\;.
\end{aligned}
\eeq
Then the time-dependent reduced density matrix of the atom can be worked out as
\begin{equation}\label{dens}
\rho(\tau)=\left(
\begin{array}{ccc}
\rho_{22}(0)e^{-\Gamma_t(\omega_0)\tau}+{\Gamma(-\omega_0)\/\Gamma_t(\omega_0)}(1-e^{-\Gamma_t(\omega_0)\tau}) & \rho_{21}(0)e^{-\Gamma_t(\omega_0)\tau/2-i\Omega\tau}\\ \rho_{12}(0)e^{-\Gamma_t(\omega_0)\tau/2+i\Omega\tau}
& \rho_{11}(0)e^{-\Gamma_t(\omega_0)\tau}+{\Gamma(\omega_0)\/\Gamma_t(\omega_0)}(1-e^{-\Gamma_t(\omega_0)\tau})
\end{array}\right)\;,
\end{equation}
where $\Omega$ is the energy level spacing of the two-level atom including the correction of the Lamb shift, and
$\Gamma_t(\omega_0)={\Gamma(\omega_0)+\Gamma(-\omega_0)}$ is the total transition rate. Eq.~(\ref{dens}) shows the effects of decoherence and dissipation on the atom. After evolving for a sufficiently long period of time $\tau\gg1/\Gamma_t(\omega_0)$, the atom will be thermalized to a steady state,
\begin{equation}\label{dens-steady}
\rho(\tau)={1\/\Gamma_t(\omega_0)}\left(
\begin{array}{ccc}
\Gamma(-\omega_0) & 0\\
0 & \Gamma(\omega_0)
\end{array}\right)\;.
\end{equation}

Now let us analyze the behavior of the transition rates, $\Gamma(\omega_0), \Gamma(-\omega_0)$, and the steady state the atom thermalizes to. To this end, we need the field correlation function ${\langle}E_i(s)E_j(0)\rangle$.
Notice that we have, for simplicity, assumed the atom to be radially polarizable, so only the component with $i=j=r$ contributes to the summation in Eq.~(\ref{rate}). The Wightman function of the radial component of the electric fields in the Unruh vacuum is given by~\cite{zhou}
\bea\label{wightman-u}
\langle 0|E_r(x)E_r(x')|0\rangle
&=&\frac{1}{4\pi}\sum_{lm}\int_{-\infty}^\infty d\omega\,\omega\,
    e^{-i\omega(t-t')}\,|\,Y_{lm}(\theta,\phi)\,|^2\nonumber\\
&&\quad\quad\quad~\times\biggl[\frac{|\,\overrightarrow{R_l}(\omega,r)\,|^2}{1-e^{-2\pi\omega/\kappa}}+
    \theta(\omega)\,|\,\overleftarrow{R_l}(\omega,r)\,|^2\biggr]\,,
\eea
where $\kappa=1/4M$ is the surface gravity of the black hole, and $\theta(\lambda)$ is the step function. So, the Fourier transform of the Wightman function (\ref{wightman-u}) with respect to the proper time is
\begin{eqnarray}\label{fourierU}
{\cal G}_{rr}(\lambda)
&=&\int^{\infty}_{-\infty}e^{i{\lambda}\Delta\tau}
   \langle 0|{E}_r(x){E}_r(x')|0\rangle\,d\Delta\tau\nonumber\\
&=&\frac{\lambda\,g_{00}}{8\pi}
   \sum_{l}(2l+1)\bigg[\theta({\lambda}\sqrt{g_{00}})|\,\overleftarrow{R_l}({\lambda}\sqrt{g_{00}},r)\,|^2
   +\frac{|\,\overrightarrow{R_l}({\lambda}\sqrt{g_{00}},r)\,|^2}{1-e^{-2\pi{\lambda}/\kappa_r}}\bigg]\;,
\end{eqnarray}
in which $\kappa_r=\kappa/\sqrt{g_{00}}\;$.
$\overrightarrow{R_l}(\omega,r)$, $\overleftarrow{R_l}(\omega,r)$ in the above equations are the radial functions of the first class of physical modes which satisfy
\beq\label{radial}
{1\/r^2}{d\/dr}\biggl[(1-2M/r){d\/dr}\big(r^2R^{(n)}_{l}(\omega,r)\big)\biggr]
+\biggl[{\omega^2\/1-2M/r}-{l(l+1)\/r^2}\biggr]R^{(n)}_{l}(\omega,r)=0\;,
\eeq
with $n=\leftarrow,\rightarrow$ labeling the modes incoming from the past null infinity and those outgoing from the past horizon respectively (see Ref.~\cite{Crispino} for details). If we rewrite the radial function as
\beq
R^{(n)}_{l}(\omega,r)=
 \frac{\sqrt{l(l+1)}}{\omega}
 \frac{\varphi^{(n)}_{\omega l}(r)}{r^2}\;,
\eeq
then Eq.~(\ref{radial}) becomes
\beq\label{raidal2}
\biggl[{d^2\/dr_*^2}+\omega^2-\biggl(1-{2M\/r}\biggr){l(l+1)\/r^2}\biggr]
\varphi^{(n)}_{\omega l}(r)=0\;,
\eeq
where $r_*=r+2M\ln(r/2M-1)$ is the Regge-Wheeler tortoise coordinate. This equation is exactly the standard spin-1 Regge-Wheeler
equation for electromagnetic perturbations of a Schwarzschild black hole.

Let us now note that the transition rates of the atom (Eq.~(\ref{rate})) can also be cast into the following form
\beq\label{Gamma}
\begin{pmatrix}
\Gamma(\omega_0)\\ \Gamma(-\omega_0)
\end{pmatrix}
=\Gamma_0(\omega_0)\bigl(\alpha_E(\omega_0)+\alpha_B(\omega_0)\bigr)
\begin{pmatrix}
1+n_\text{eff}(\omega_0)\\n_\text{eff}(\omega_0)
\end{pmatrix}\;,
\eeq
where $\Gamma_0(\omega_0)=\omega_0^3\,|\langle 1|{\bf D}|2\rangle|^2/3\pi$ is the spontaneous emission rate in flat spacetime, and we have defined
\bea\label{alphaE}
\alpha_E(\omega_0)
&=&{3g_{00}\/8\omega_0^2}\sum_{l}(2l+1)
  |\,\overleftarrow{R_l}(\omega_0\sqrt{g_{00}},r)\,|^2\;,\\
\alpha_B(\omega_0)
&=&{3g_{00}\/8\omega_0^2}\sum_{l}(2l+1)
  |\,\overrightarrow{R_l}(\omega_0\sqrt{g_{00}},r)\,|^2\;,
\label{alphaB}
\eea
which are associated with the incoming (environment) and outgoing (black hole) field modes, and an effective photon number
\beq\label{neff}
n_\text{eff}(\omega_0)=
 \frac{\alpha_{B}(\omega_0)\,n(\omega_0,\kappa_r/2\pi)}
      {\alpha_{E}(\omega_0)+\alpha_{B}(\omega_0)}\;,
\eeq
with $n(\omega,T)=(e^{\omega/T}-1)^{-1}$.
In the two asymptotic regions, we have
\bea\label{}
&r\rightarrow2M\,:\left\{
\begin{aligned}
\alpha_E(\omega_0)\approx
 &{3\/8\omega_0^4r^4}\sum_{l}l(l+1)(2l+1)|\mathcal{T}_l(\omega_0\sqrt{g_{00}})|^2\;,\\
\alpha_B(\omega_0)\approx
 &{3\/8\omega_0^4r^4}\sum_{l}l(l+1)(2l+1)
   \big[\,1+|\overrightarrow{\mathcal{R}_l}(\omega_0\sqrt{g_{00}})|^2\\
  &+\overrightarrow{\mathcal{R}_l}(\omega_0\sqrt{g_{00}})e^{-2i\omega_0\sqrt{g_{00}}r_*}
   +\overrightarrow{\mathcal{R}_l}^*(\omega_0\sqrt{g_{00}})e^{2i\omega_0\sqrt{g_{00}}r_*}
  \big]\;,
\end{aligned}\right.\\
&r\rightarrow\infty\,:\left\{
\begin{aligned}
\alpha_E(\omega_0)\approx
 &{3\/8\omega_0^4r^4}\sum_{l}l(l+1)(2l+1)
  \big[\,1+|\overleftarrow{\mathcal{R}_l}(\omega_0\sqrt{g_{00}})|^2\\
  &+\overleftarrow{\mathcal{R}_l}(\omega_0\sqrt{g_{00}})e^{2i\omega_0\sqrt{g_{00}}r_*}
   +\overleftarrow{\mathcal{R}_l}^*(\omega_0\sqrt{g_{00}})e^{-2i\omega_0\sqrt{g_{00}}r_*}
  \big]\;,\\
\alpha_B(\omega_0)\approx
 &{3\/8\omega_0^4r^4}\sum_{l}l(l+1)(2l+1)|\mathcal{T}_l(\omega_0\sqrt{g_{00}})|^2\;.
\end{aligned}\right.
\eea
So, the transition rates of the atom are position-dependent since $\alpha$ and $n_\text{eff}$ are both functions of $r$ and this variation with the position is a reflection of the physical characters of the black hole. To be specific, the transition rates are determined by the coefficients $\alpha_E(\omega_0)$, $\alpha_B(\omega_0)$ reflecting how propagation of the field modes is affected by  the  curved geometry of the spacetime, and the  photon number $n(\omega_0,\kappa_r/2\pi)$ characterizing the Hawking radiation of the black hole. The result takes the same form as that of a two-level atom in a stationary environment out of thermal equilibrium formed by a body of arbitrary geometry and dielectric permittivity at temperature $T_M$ surrounded by walls at a different temperature $T_W$~\cite{Bellomo13}, where the coefficients $\alpha_W(\omega_0)$, $\alpha_M(\omega_0)$ determined by the reflection and transmission scattering operators exhibit the modification of the fields caused by the body, and the effective photon number $n_{\rm eff}(\omega_0)$ is a weighted average of the radiation contributed by the body, as well as the surrounding walls. So the transition rates of the atom are modified by the presence of a material body because of the changes of the local fields caused both by the emission of the fields and scattering of those coming from the environment by the body. So our case is analogous  to the situation of a zero wall temperature, where the effects of the Hawking radiation and the spacetime backscattering of vacuum electromagnetic fields are mimicked by the thermal radiation and the scattering of the fields caused by the material body.

Using the properties of the radial functions (Eqs.~(59) and (60) of Ref.~\cite{zhou}), one can show that  $\alpha_{E}(\omega_0)$, $\alpha_{B}(\omega_0)$, i.e.,  Eqs.~(\ref{alphaE}), (\ref{alphaB}), take the following approximate forms in the asymptotic regions
\beq\label{asymp-alpha}
\left\{
\begin{aligned}
&r\rightarrow2M\,: &&\alpha_E(\omega_0)\approx f(\omega_0,2M)\,,\quad
&&\alpha_B(\omega_0)\approx1+{a^2(r)\/\omega_0^2}\,,\\
&r\rightarrow\infty\,:
&&\alpha_E(\omega_0)\approx1\,,\quad
&&\alpha_B(\omega_0)\approx f(\omega_0,r)\,,
\end{aligned}\right.
\eeq
where $f(\omega,r)$ is given by
\beq\label{grey}
f(\omega,r)=
 \frac{3}{8r^4\omega^4}\sum_ll(l+1)(2l+1)\;|\mathcal{T}_l(\omega\sqrt{g_{00}})|^2\;,
\eeq
and
\beq
a(r)=\frac{M}{r^2\sqrt{g_{00}}}=\frac{M}{r^2\sqrt{1-2M/r}}\;
\eeq
is the proper acceleration of the static atom at position $r$.
With the help of Eqs.~(\ref{Gamma}) and (\ref{asymp-alpha}), we obtain, in the two asymptotic regions, the transition rates of the atom
\beq\label{}
\Gamma(\omega_0)\approx\left\{
\begin{aligned}
  &\Gamma_0(\omega_0)\bigg[\bigg(1+{1\/e^{2\pi\omega_0/\kappa_r}-1}\bigg)\bigg(1+{a^2(r)\/\omega_0^2}\bigg)
  +f(\omega_0,2M)\bigg]\;,
  &r\rightarrow2M\;,\\
  &\Gamma_0(\omega_0)\bigg[1+\bigg(1+{1\/e^{2\pi\omega_0/\kappa_r}-1}\bigg)f(\omega_0,r)\bigg]\;,
  &r\rightarrow\infty\;,
\end{aligned} \right.
\eeq
\beq\label{}
\Gamma(-\omega_0)\approx\left\{
\begin{aligned}
  &\frac{\Gamma_0(\omega_0)}{e^{2\pi\omega_0/\kappa_r}-1}\bigg(1+{a^2(r)\/\omega_0^2}\bigg)\;,
  &r\rightarrow2M\;,\\
  &\frac{\Gamma_0(\omega_0)}{e^{2\pi\omega_0/\kappa_r}-1}\,f(\omega_0,r)\;,
  &r\rightarrow\infty\;.
\end{aligned} \right.
\eeq
The appearance of a Planckian factor in the transition rates suggests that there is thermal radiation from the black hole, and it is this thermal radiation that leads to a nonzero spontaneous excitation rate $\Gamma(-\omega_0)$. In the vicinity of the horizon, there is an additional term proportional to $a^2$ in the transition rates as compared to the scalar field case~\cite{yu1}. Such a term is also present in the spontaneous excitation~\cite{ZYL06} and energy level shifts~\cite{Passante} of an accelerated atom coupled with electromagnetic fluctuations in the Minkowski vacuum. Far away from the black hole,  the thermal terms in the transition rates are modified by a grey-body factor $f(\omega_0,r)$ which vanishes at spatial infinity.  This indicates that the thermal flux is backscattered by the spacetime curvature. The grey-body factor also exists in the spontaneous emission rate near the horizon, which is the contribution of the vacuum fluctuations of the ingoing modes. However, such a term is absent in the spontaneous excitation rate since  it cancels out with the corresponding radiation reaction parts~\cite{zhou}.

As the spontaneous transition process persists for a sufficiently long time ($\tau\gg\Gamma_t^{-1}$), the ratio of population of the atoms in the ground and excited states reaches a steady value, and the equilibrium state of the atom Eq.(\ref{dens-steady}) can be described as a mixed thermal state
\beq
\rho_{\infty}=
 {e^{-H_A/T_\text{eff}} \/ {\rm Tr}[e^{-H_A/T_\text{eff}}]}\;,
\eeq
with an effective temperature
$T_\text{eff}(\omega_0)=
 \omega_0\big/\ln\bigl(1+1/n_\text{eff}(\omega_0)\bigr).$
Therefore, regardless of its initial state, a static two-level atom outside the black hole is asymptotically driven to a thermal state at temperature $T_\text{eff}$. Remarkably, here the thermalization temperature is dependent on the transition frequency $\omega_0$ of the atom, which implies that atoms with different transition frequencies thermalize to different temperatures. This may seem surprising and pose a conceptual challenge to our common physical understanding  at first glance.  It is, however, in close analogy to what happens for a two-level atom in a stationary environment out of thermal equilibrium recently studied in Ref.~\cite{Bellomo13}. This is a reflection of the fact that the situation we are considering is a black hole radiating into empty space, which exhibits a nonequilibrium nature.

Now we examine how the effective temperature $T_\text{eff}(\omega_0)$ varies with the radial position $r$. It can be inferred from Eqs.~(\ref{neff}), (\ref{asymp-alpha}) that the effective temperature is related to the grey-body factor $f(\omega,r)$ in the asymptotic regions. Although the explicit form is unknown, we can analyze its behaviors in the limit of low and high frequencies. At low frequencies $M\omega\ll1$, the transmission coefficient takes the following form: $|\mathcal{T}_l(\omega)|^2\approx 4\big[{{(l+1)!(l-1)!}\/{(2l)!(2l+1)!!}}\big]^2(2M\omega)^{2l+2}$~\cite{Fabbri}, then $f(\omega,r) \propto g_{00}^2\,M^4/r^4$. At high frequencies $M\omega\gg1$, we can apply the geometrical optics approximation $\mathcal{T}_l(\omega\sqrt{g_{00}})\sim \theta(\sqrt{27}M\omega\sqrt{g_{00}}-l)$~\cite{DeWitt}; thus we also have $f(\omega,r) \propto g_{00}^2\,M^4/r^4$. In both cases, only the leading term is kept in the summation. Here let us note that, since the frequency appearing in the transmission coefficient is multiplied by a factor $\sqrt{g_{00}}$ in the definition of the grey-body factor Eq.~(\ref{grey}), the low frequency condition $M\omega\ll1$ will always be fulfilled for  atoms  placed close enough to the horizon. Meanwhile, for a given frequency, the grey-body factor always tends to zero as long as the atom is placed far enough from the black hole, since it decays with $r$ as $1/r^4$. So, we conclude that the grey-body factor $f(\omega,r)$ vanishes in both two asymptotic regions, and is independent of the transition frequency $\omega_0$. Therefore, for a given atom, the effective temperature tends to $\kappa_r/2\pi$ when the atom is placed close to the horizon, and to zero when the atom is placed infinitely far from the black hole. For an arbitrary position $r$, we can see from Eqs.~(\ref{neff}) and the expression for $T_\text{eff}(\omega_0)$ that the effective temperature of the atom always lies between zero and $\kappa_r/2\pi$, and moreover it depends on the transition frequency $\omega_0$. The situation is similar to that of an atom in front of a dielectric slab in a stationary environment out of thermal equilibrium~\cite{Bellomo13}. When the atom is placed close to the slab, the effective temperature tends to the temperature of the slab. For the atom placed far from the slab, the effective temperature is generally dependent on the temperature, and the geometric and dielectric properties of the slab. However, if the slab is free of energy dissipation, i.e., its permittivity is real, then the effective temperature tends to the temperature of the environment. So, the dynamical behaviors of a two-level atom outside a radiating black hole, such as its lifetime and the steady state it thermalizes to, closely resemble those near a dielectric body with desired real permittivity in a stationary environment out of thermal equilibrium, and this in principle opens up a possibility of verifying Hawking radiation of black holes in experiment, using artificially engineered dielectric materials.

Let us now discuss in detail the possibility of testing our results in a simulated fashion. One may choose to test the behaviors of atoms outside a black hole at a radial position $r\geq r_0$. Here $r_0$ denotes a position close to the horizon. A dielectric body kept at a temperature $\kappa_{r_0}/2\pi$ in vacuum with its ambient temperature close to zero forms the desired analogue stationary environment out of thermal equilibrium. 
With the geometry and dielectric permittivity appropriately designed, the peculiar transition frequency dependence of the dynamical behaviors of two-level atoms at a radial distance $r\geq r_0$ outside the black hole can then be simulated by those at a position $z\geq z_0$ from the dielectric body. Superconducting qubits  can be applied as frequency-tunable artificial two-level atoms, since unlike real atoms, their characteristics can be tailored at will. In practice, the designing of such metamaterials may not be an easy task since it is difficult to deduce both the geometric and dielectric properties  required to generate dynamical behaviors of a nearby atom which are the same as that of an atom outside a radiating black hole. However, the deduction of the dielectric properties for a body with certain geometry may be possible. For example, in the case of a slab as studied in Ref.~\cite{Bellomo13}, the coefficients $\alpha_M$, $\alpha_W$ which directly relate to  the dynamics of the atom can be derived as functions of the dielectric permittivity $\varepsilon(\omega)$.  Therefore, the behaviors of atoms with different transition frequencies at a certain distance outside a black hole can be demonstrated by those in front of a dielectric body of certain geometry, e.g., a slab, made of metamaterials with a required dispersion relation $\varepsilon(\omega)$.

In summary, we have studied the dynamical evolution of a radially polarizable two-level atom coupled with quantum vacuum electromagnetic fields at a fixed radial distance outside a radiating Schwarzschild black hole, and calculated the atomic transition rates and the effective temperature of the steady state. We have found that, regardless of its initial state, the atom will be asymptotically driven to a thermal state at an effective temperature which is dependent on the transition frequency of the atom. The dynamical behaviors of the atom, such as its lifetimes and the thermalization process, are similar to those near a dielectric body in a stationary environment out of thermal equilibrium.  Our results therefore suggest, in principle, a possibility to verify the peculiar features of the Hawking radiation by observing the dynamical behaviors we find here for a two-level atom in tabletop experiments using engineered materials with desired dielectric properties and superconducting circuits for an experimental implementation of two-level atoms.


This work was supported in part by the NNSFC under Grants No. 11075083, No. 10935013, and No. 11375092; the
Zhejiang Provincial Natural Science Foundation of China under Grant
No. Z6100077; the
National Basic Research Program of China under Grant No.2010CB832803; the PCSIRT under Grant No. IRT0964;
the Hunan
Provincial Natural Science Foundation of China under Grant No.
11JJ7001; the SRFDP under Grant
No. 20124306110001; and Hunan Provincial Innovation Foundation for
Postgraduate under Grant No. CX2012A009.


\end{document}